\documentclass{camera}
\usepackage{epsfig}
\begin{document}
\begin{picture}(1,1)
\put(-10,20){
\begin{minipage}{10cm}
{\small
``Nanotubes and Nanostructures''        \\
S. Bellucci (Ed.)                       \\
SIF, Bologna, 2001                      \\
ISBN: 88-7794-291-6                     \\
}
\end{minipage}
}
\end{picture}

\vspace*{.1cm}

\title{Transport through a \\ one--dimensional quantum dot}

\author{T.\ Kleimann$^{a}$, M.\ Sassetti$^{a}$, G.\ Cuniberti$^{b}$, \And B.\ Kramer$^{c}$}

\organization{$^{a}$Dipartimento di Fisica, INFM, Universit{\`a} di
  Genova,\\ Via Dodecaneso 33, I--16146 Genova, Italy\\
$^{b}$Max-Planck-Institut f{\"u}r Physik komplexer Systeme,\\
N{\"o}thnitzer Stra{\ss}e 38, D--01187 Dresden, Germany\\
 $^{c}$ I. Institut
  f\"ur Theoretische Physik, Universit\"at Hamburg, Jungiusstra{\ss}e
  9, D--20355 Hamburg, Germany}

\maketitle \abstract{We examine the effects of long-range interactions
  in a quantum wire with two impurities. We employ the bosonization
  technique and derive an effective action for the system. The effect
  of the long-range interaction on the charging energy and spectral
  properties of the island formed by the impurities and the linear
  transport is discussed.}
\section{Introduction}
Considerable progress in the fabrication of quasi one-dimensional
nanostructures has been made in recent years. Such quantum wires are
ideal tools for studying the interplay of impurities and interaction
effects. It is even possible to form one--dimensional quantum dots
\cite{aus99} and investigate their transport properties in the linear
and nonlinear regime where the latter gives insight into the spectral
properties of the system. In quantum wires fabricated by the
cleaved-edge-overgrowth technique \cite{yac97}, the electron density
is decreased by an external gate voltage above the wire region.
Eventually, the density becomes so low that impurity potentials
traverse the Fermi energy, and a one--dimensional quantum island is
formed between two such impurities. Since we can treat the low energy
properties of one--dimensional interacting electrons in terms of
Luttinger liquid theory, the role of many body effects can be
investigated. In particular, the interfacing between the quantum dot
and the leads is technologically important and byproducts of the
effort in understanding transport through such nanostructures may lead
to operational criteria for future devices.
\section{Model}
We use the bosonization technique \cite{hal,voit} for interacting
electrons in 1D. The quantum dot is described by a double
barrier consisting of two delta-function potentials $V_{i}\,\delta
(x-x_{i})$ at $x_i\, (i=1,2)$ where $x_1<x_2$. An external electric
field ${\cal E}(x,t)=-\partial_{x} U(x,t)$ is assumed to induce
transport. The Hamiltonian is
\begin{equation}
  \label{hamiltonian}
 H= H_{0} + H_{1} + H_{2}\,.  
\end{equation}
The first term describes the interacting electrons as a Luttinger liquid
\cite{tom,lutt},
\begin{eqnarray}
\label{eq:1}
 H_{0} &=& 
\frac{\hbar v_{\rm F}}{2}\int \!\! {\rm d}x \; \Big[\Pi^{2}(x)  + [\partial_{x}\vartheta(x)]^{2}\Big] + \frac{1}{\pi}\int \!\!{\rm d}x \! \! \int \!\!{\rm d}x'\, \partial_{x}\vartheta(x)\,V(x-x')\,
\partial_{x'}\vartheta(x')\nonumber\\
 &&\nonumber\\
\end{eqnarray}
The electrons are represented by conjugate {\sl bosonic plasmon
  fields} $\Pi(x,t)$ and $\vartheta(x,t)$ which are associated with the
collective charge density excitations.  The system length is $L$ and
the Fermi velocity $v_{\rm F}$. The interaction, $V(x-x')$, is a
projection of a modified 3D Coulomb interaction onto the $x$-direction
(see below). It has the Fourier transform $\hat{V}(q)$, which is the
dominant quantity in the dispersion relation of the collective charge
excitations
\begin{equation}
  \label{eq:2}
\omega(q) = v_{\rm F}
\left|q\right|\left[ 1+ \frac{\hat{V}(q)}{\pi\hbar v_{\rm F}} \right]^{1/2}.
\end{equation}
The dispersion relation for free electrons would be linear
($\omega(q)=v_{\rm F}|q|$). The transport inducing electric field
(slowly varying with respect to $k_{\rm F}^{-1}$) couples to the long
wavelength parts of the charge density $\rho(x) \approx \rho_0+
(1/\pi)^{1/2}\,\partial_{x} \vartheta(x)$. Here $\rho_0=k_{\rm
  F}/\pi$ is the mean electron density. This yields
\begin{equation}
H_{2} = -e \sqrt{\frac{2}{\pi}} \,
       \int{\rm d}x \, U(x,t) 
       \partial_{x}\vartheta(x).
\label{drive}
\end{equation}
Finally, the two localized impurities define the left and right lead
and the quantum dot itself between the impurities. Their contribution
is
\begin{eqnarray}
H_{\rm 2} = \rho_0\sum_{i=1,2} V_i \cos{[2k_{\rm F}x_{i}+
  2 \sqrt{\pi}\vartheta(x_{i})]} 
  \label{barrier}
\end{eqnarray}
Because of gates and surrounding charges in the actual experiments,
screening is always present for the mutual electron interaction. We
model this situation in terms of a 3D Coulomb interaction, an infinite
metallic gate in distance $D$ to the quantum wire, and parabolic
confinement perpendicular to the wire. The effective interaction
potential is then obtained by using the method of image charges and
projecting onto the wire axis with a normalized Gaussian confinement
wavefunction. This leads to the Fourier transform of the effective
interaction potential \cite{klm}
\begin{eqnarray}
\hat{V}(q)= 
V_0\left[\, {\rm e}^{d^2q^2/4} \,
{\rm E}_{1} \left(\frac{d^2q^2}{4}\right) - 
2{\rm K}_{0}(2Dq)\right]\,,
\label{1dqpot}  
\end{eqnarray}
where $V_{0}=e^{2}/4\pi\epsilon_{0}\epsilon$,
$\varepsilon_{0}\epsilon$: dielectric constant, $D\gg d$: gate
distance and wire diameter, E$_{1}(z)$: exponential integral, and
${\rm K}_{0}(z)$: modified Bessel function \cite{abramowitz}.
\section{Effective Action}

For determining the transport properties through the system it is
sufficient to know the behavior of the plasmon fields at the barrier
positions $x_1$ and $x_2$. Hence, we calculate an effective action for
the symmetric and antisymmetric variables for the density
\begin{equation}
\label{variable}
N^{\pm} = \frac{1}{\sqrt{\pi}}\Big[\vartheta(x_{2}) \pm
\vartheta(x_{1})\Big]\,,
\end{equation}
which are related to the excess particle number on the island between
the two impurities ($-$) with respect to the mean value. The $+$-sign
refers to the number of im--balanced particles between the left and
right lead. The effective action is evaluated by use of an
imaginary-time path integral method \cite{sk} and reads
\begin{eqnarray}
S_{\rm eff}[N^{+},N^{-}] &=& \int_{0}^{\hbar \beta}\! \!  {\rm d}\tau \; H_{2}[N^{+},N^{-}] - \sum_{r=\pm} N^{r}(\tau) {\cal L }^{r} (\tau) \nonumber\\
& & +\frac{1}{2}\sum_{r=\pm}  \int_{0}^{\hbar \beta}\! \! \int_{0}^{\hbar \beta} \! \! {\rm d}\tau {\rm d}\tau' \;
N^{r}(\tau) K^{r}(\tau -\tau') 
N^{r}(\tau').
\label{effaction}
\end{eqnarray}
The dissipative kernels $K^{\pm}(\tau)$ and the effective driving
force ${\cal L}^{\pm}(\tau)$ capture the effects of the interaction
through the dispersion relations (\ref{eq:2}) of the collective modes.
While ${\cal L}^{\pm}(\tau)$ describes the effect on the external
field and is considered elsewhere \cite{klm}, $K^{\pm}(\tau)$ is
associated with the charging energy and the spectral properties of the
island. It has the Fourier transform at Matsubara frequencies ($L
\rightarrow \infty$)
\begin{eqnarray}
\left[K^{\pm} (\omega_{n})\right]^{-1}= 
\frac{4 v_{\rm F}}{\hbar\pi^2} \int_{0}^{\infty}\!{\rm d}q  
\frac{1 \pm \cos [q(x_{1} -x_{2})]}{ \omega_{n}^2 +
\omega_{\nu}^2(q)}.
\label{kernel}
\end{eqnarray}
\section{Results}
\begin{figure}[htb]
\begin{center}
\setlength{\unitlength}{1.0cm}
\epsfig{file=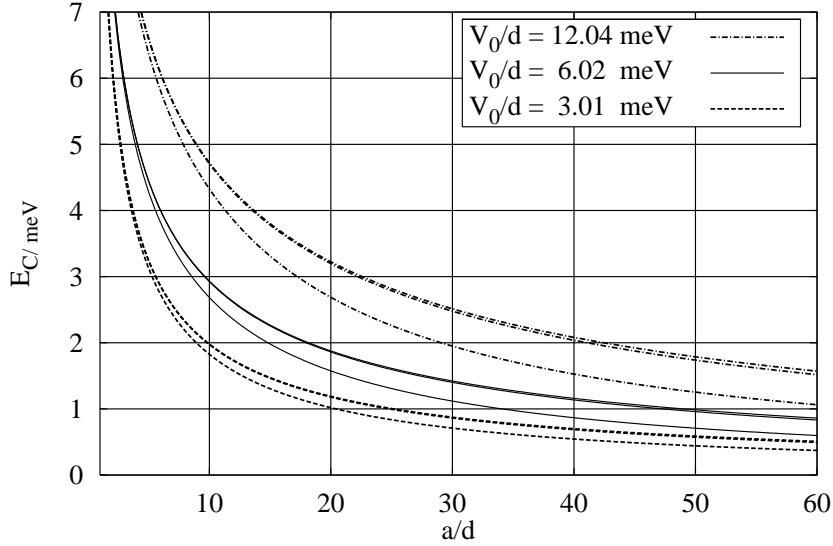,width=11cm}
\parbox{13cm}{\caption{\label{figcharg}Charging energy $E_{\rho}$ in meV as a
    function of $a/d$ (Fermi energy $E_{\rm F}=3$\,meV, effective
    electron mass $m=0.067m_{0}$), $d=20$nm, interaction strengths
    $V_{0}/d=12.04$\,meV (dashed-dotted), 6.02\,meV (full lines,) and
    3.01\,meV (dotted), corresponding to $\epsilon = 3$, 6, 12,
    $D/d=500$, 50, 5 (top to bottom).}}
\end{center}
\end{figure}

At low temperatures the transport through the quantum island is
dominated by charging effects. The {\em linear conductance} shows
discrete maxima which correspond to the transfer of single electrons
through the dot. The peaks occur whenever the chemical potentials of
the leads and the quantum dot are aligned. Otherwise, transport is
blocked (Coulomb blockade). By applying a gate voltage to the island,
the situation of waved Coulomb blockade can be achieved periodically
and one obtains so--called Coulomb blockade oscillations. The distance
between the conductance peaks with respect to the gate voltage is then
directly related to the charging energy of the dot and experimentally
accessible. From equation (\ref{kernel}) an analytic, microscopic
expression for the charging energy is readily obtained in the static
limit: $E_{\rm C}= K^{-}(\omega_n \to 0)/2$.  For free electrons, the
charging energy is only due to the Pauli exclusion principle and can
be evaluated analytically: $E_{\rm free}=\pi\hbar v_{\rm F}/2a$.
Generally, the charging energy depends on the system parameters
through the dispersion in equations (\ref{eq:2}) and (\ref{1dqpot}).
The parameters are interaction strength $V_0$, interaction range which
is basically determined by the distance of the gate $D$, the length of
the island $a$, and the wire diameter $d$. Fig. \ref{figcharg} shows
the numerically evaluated charging energy for realistic experimental
values \cite{aus99}.
We find that the charging energy increases with increasing interaction
range $D$. For a very large $D$, an asymptotic value is reached. This
is because in the beginning, the increasing range makes it harder for
electrons to get on the island.  But increasing the range even further
by a large $D$ with respect to the island length $a$ does not change
the charging energy significantly because the repulsion between the
electrons {\sl on} the island is the dominant contribution. Certainly
the charging energy increases when the overall interaction strength
$V_0$ is increased. We find consistent values for charging energy and
island length for the experimental parameters of Ref.\ \cite{aus99}
within our microscopic approach.

\begin{figure}[htb]
\begin{center}
  \epsfig{file=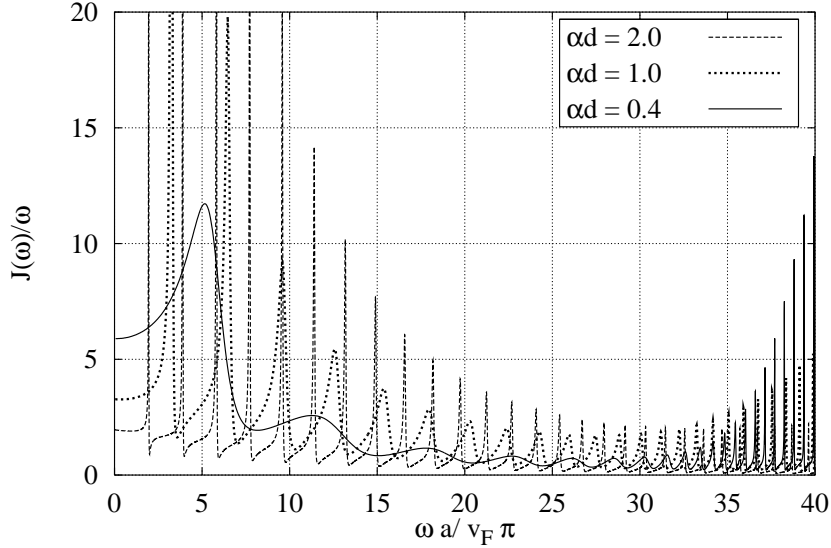,width=11cm}
  \parbox{13cm}{\caption{\label{figspec}Spectral densities
      $J(\omega)/\omega$ for the interaction $V(x)= 2V_0/(\alpha d)^2
      \exp{(-\alpha |x|)}$ as a function of normalized frequency
      $\omega a/ v_{\rm F} \pi$ where $a/d=10\pi$ and
      $g_0=[1+V_0/(\hbar \pi v_{\rm F})]^{-1/2}=0.5$. The interaction
      range is tuned by $\alpha d=2$, 1, 0.4 (increasing range).}}
\end{center}
\end{figure}

The frequency-dependent parts of the kernels describe the dynamic
effects of the external leads and of the correlated excited states in
the dot.  Their influence is described by spectral densities
$J^{\pm}_{\nu}(\omega)$ which are related to the imaginary-time
kernels via analytic continuation.
\begin{equation}
  \label{spectral}
J(\omega)=
\frac{1}{\pi\hbar} \sum_{\nu=\pm} {\rm Im} K^{\nu}(\omega_n\to+{\rm i}\omega)\,.
\end{equation}
For our realistic long ranged interaction (\ref{1dqpot}) analytic
expressions for these densities are not available, but we can learn
about their properties by using a ''long range toy'' interaction
$V^{\rm toy}(x)= 2V_0/(\alpha d)^2 \exp{(-\alpha |x|)}$ where $\alpha$
is a phenomenological screening parameter which determines the range
of the interaction. The relatively simple form of the Fourier
transform of this interaction allows for the analytical calculation of
the nonlocal conductivity of a pure quantum wire \cite{cun}. The
kernel (\ref{kernel}) can also be written in terms of the
conductivity.  Lengthy but straightforward calculations finally yield
the spectral density.  This spectral density is plotted in Fig.\ 
\ref{figspec} for three different $\alpha$.
It displays the following features due to the long range interaction:
there are broadened, delta-function-like peaks which tend to become
equidistantly spaced for large $\omega$. The peaks become very narrow
for large energies. This effect is enhanced when the range of the
interaction is decreased. Actually the limit $\alpha \to \infty$ leads
to non-interacting electrons as seen in the form of $V^{\rm toy}$.
Further, the peaks are strongly broadened when the interaction range
is increased. Again, for a large $\omega$, independent of the value of
$\alpha$, the peaks become narrow, almost delta-function-like peaks,
and attain an equidistant spacing which we would expect for
non-interacting electrons in one dimension which are confined to a
region of length $a$. We suggest the following interpretation: the
interacting electrons are described by their collective excitations in
terms of the plasmon fields. The barriers pin these plasmons and a
situation similar to a standing wave occurs.  The excitation energies
of the system then approximately correspond to $\omega(q_m=m \pi/a)$.
This explains why the maxima are shifted when the interaction range is
increased. It is due to the nonlinearity of the dispersion relation
for the plasmons at ``smaller energies'' while the dispersion becomes
linear again after a certain energy $\omega_{\rm p}$ \cite{cun}. On
the other hand, the effective Hamiltonian is equivalent to one for a
quantum Brownian particle (corresponding to the coordinates $N^\pm$)
in a dissipative environment and external potential. In this context
the spectral density describes the possible ``channels'' for energy
dissipation between the environment and the Brownian particle in a
potential. One finds in the zero-range case that the peak positions
correspond to (single) particle states in the dot. This analogy leads
us to interpreting the peaks as states of the quantum dot which may
contribute to transport through the dot.  Because the interaction
couples the dot with the leads, the ``levels'' of the quantum dot are
broadened. The leads contribute mainly to the spectral density through
a quasi Ohmic background. In the case of free electrons, the spectral
density can be separated into a sum of delta function peaks describing
the states of the dot and an Ohmic part ($\propto \omega$)
corresponding to the leads. This separation also works if one takes a
zero-range interaction. Basically this tells us that a zero-range
interaction creates an isolated quantum dot coupled to leads. Taking
into account a long-range interaction, accounts for the correlation
effects better, and the corresponding spectral density describes the
overall spectral properties of the entire system including the left
lead, the dot and the right lead. To sum up, first the interaction
shifts the energy levels of the dot, and second the levels are
broadened due to the coupling to the leads.

Though we are unable to calculate the complete spectral densities for
our realistic interaction (\ref{1dqpot}) we can obtain the limits for
$\omega\to 0$,
\begin{equation}
J(\omega\to 0)= \frac{\omega}{2g}\left[ 1+  g^4 \left(\frac{2 a E_{\rm C}}{\pi \hbar v_{\rm F}} \right)^2 \right],
\label{speclimit}
\end{equation}
where $g^{-2}=1+\hat{V}(q\rightarrow 0)/\pi \hbar v_{\rm F}$ and
$E_{\rm C}$ is the charging energy. This limit describes the
dissipative influence of the low-frequency charge excitations in the
external leads, $x < x_{1}$ and $x > x_{2}$. It holds also for finite
frequencies. However, these must be smaller than the frequency scale
corresponding to the range of the interaction, and smaller than the
characteristic excitation energy of the correlated electrons in the
dot. One can show that the intrinsic width $\Gamma(T)$ of the
conductance peaks of the Coulomb oscillations exhibit a nonanalytic
power-law due to the interaction \cite{furu93,braggio}. The exponent
is directly related to the limit in eq.\ (\ref{speclimit}). Defining
$1/g_{\rm eff}:=J(\omega\to 0)/\omega$ we write for the temperature
dependence $\Gamma(T)\propto T^{1/g_{\rm eff}-1}$ \cite{klm}. In
contrast to a treatment with zero-range interaction, the exponent
depends in a nontrivial way on the geometry of the system and the
parameters of the interaction reflected by the charging energy $E_{\rm
  C}(a, D, V_0)$ in eq.\ (\ref{speclimit}). In the zero-range case
$g_{\rm eff}$ becomes simply equal to $g$ and the coupling of the dot
seems to be disguised -- only the properties of the leads are manifest
in terms of $g$. Thus the long range nature of the interaction should
be manifest in the shape of the Coulomb peaks in experiments.
\section{Conclusion}
We derived an effective action for a double barrier in a quantum wire
taking account the interactions in terms of Luttinger liquid theory.
We discussed the effect of a realistic, long-range interaction on the
charging energy which relates to the distances between Coulomb
blockade peaks. The spectral properties of the system are examined by
resorting to a simpler ``toy interaction'' which shows the basic
features of long range interaction. The spectral density of the system
is compared to the case of the usual zero range Luttinger liquid
interaction. Finally, the effect of the interaction on the shape of
the linear conductance Coulomb peaks is mentioned.

This contribution only highlights the role of the charge degree of
freedom. One can also include effects due to the electron spin
\cite{klm}. In the bosonized picture, the degrees of freedom are
decoupled and develop charge and spin collective excitations
respectively. In addition to the results above spin addition energies
and spin excited states appear. Also, the power law for intrinsic
width of the Coulomb peaks acquires a spin dependence.

\end{document}